# Window Function-less DFT with Reduced Noise and Latency for Real-Time Music Analysis

Cai Biesinger
Graduate School of Informatics
Kyoto University
Kyoto, Japan
cai.biesinger.78m@st.kyoto-u.ac.jp

Hiromitsu Awano
Graduate School of Informatics
Kyoto University
Kyoto, Japan
awano@i.kyoto-u.ac.jp

Masanori Hashimoto
Graduate School of Informatics
Kyoto University
Kyoto, Japan
hashimoto@i.kyoto-u.ac.jp

*Abstract*—Music analysis applications demand algorithms that can provide both high time and frequency resolution while minimizing noise in an already-noisy signal. Real-time analysis additionally demands low latency and low computational requirements. We propose a DFT-based algorithm that accomplishes all these requirements by extending a method that post-processes DFT output without the use of window functions. Our approach yields greatly reduced sidelobes and noise, and improves time resolution without sacrificing frequency resolution. We use exponentially spaced output bins which directly map to notes in music. The resulting improved performance, compared to existing FFT and DFT-based approaches, creates possibilities for improved real-time visualizations, and contributes to improved analysis quality in other applications such as automatic transcription.

## I. Introduction

Audio analysis applications like song recognition, speech analysis, music information retrieval, etc. are widely used for education, training, and entertainment. Some applications require real-time music analysis, such as singing evaluation and music visualization. Music visualization, by stimulating more than one sense of the listener at once, allows for a more engaging, fun, and immersive listening experience both in private and at live events, and acts as an additional form of artistic expression. It is also used as an educational tool, helping further the understanding of the structure and details of a piece of music [1].

The Fast Fourier Transform (FFT) is an extremely common tool used in a variety of audio analysis applications. However, it has several shortcomings which make it sub-optimal for real-time music analysis. This use case demands low latency, low computational complexity, and output in a form suitable for the analysis of music. Low latency is especially imperative for music visualization applications. A visualization becomes pointless if latency is too high, as correlation with the music, and therefore immersion, is lost. Low computational complexity is mandatory to run applications in real time even on low-performance computers. Low noise and high specificity are also desirable to minimize the influence of noise content from other instruments, distortion, percussion, etc. [2]

We propose an improved algorithm in this work simultaneously achieving low latency, computational complexity, and noise. We developed it for use in a real-time music visualization system that is responsive, relatable, and broadly applicable. The proposed algorithm exploits the characteristics of music visualization in that an inverse transform is not required, and slight inaccuracies are tolerable. We also believe that the algorithm is more broadly applicable to music analysis tasks, many of which begin by applying a DFT to the input signal. By using an algorithm which creates a more suitable frequency-space representation, such analysis may also achieve enhanced performance and output quality. For example, transcription accuracy of automatic music transcription may improve with our approach, since it is an application that benefits greatly from lower noise and higher resolution analysis [3].

Our work is based on the *neighbour spectrum sample components composition* (NC) method [4], which post-processes FFT output to remove sidelobes and increase frequency resolution without the use of a window function. However, the authors did not provide any insight into why and how the NC method functions, nor an explanation of the conditions necessary for it to operate, meaning it could only be applied to FFTs. Our work in this paper generalizes the NC method to be usable outside of FFTs, proposes a novel DFT approach that is highly tailored for music processing, and explains how such a DFT can be implemented in a computationally efficient manner. Our method maintains the advantages that sidelobes are removed, and frequency resolution is doubled without the use of a window function. We tailored the NC method to music visualization and implemented a software system to process music containing polyphonic melodies in real time.

This paper is organized as follows. Necessary key ideas are introduced in Section II, then we step through our generalization and characterization of the NC method in Section III. Section IV explains our proposed DFT implementation and provides motivations for our choices. Section V explains the practical results and our implementation's performance. Concluding remarks are given in Section VI.

## II. Background

The common Western equal temperament half-tone music scale is represented by a series of 12 notes per octave with the following exponential relationship for a pitch $n$:

$$f(n) = 440\text{Hz} \times 2^{(n-69)/12}. \quad (1)$$

The required window size of a regular Discrete Fourier Transform (DFT), like the FFT, can be calculated from the desired output bin spacing $\Delta f$ at sample rate $F_S$ with sampling theorem:

$$N = \frac{F_S}{2\Delta f}. \quad (2)$$

The FFT and many other DFTs output a linear series of frequency bins, which is unsuitable for music analysis. Instead, an exponential series mirroring (1) is desired. Additionally, the output resolution of an FFT is directly correlated with the input window size. For example, to distinguish A0 (27.5Hz) and A♯0 (29.14Hz) notes by 1 bin at a 48KHz sample rate would require a window size of minimum 14,634 samples (0.305s). However, this resolution becomes detrimental in high frequencies, shown by the range of just the A6 note (1710.61Hz to 1812.33Hz) spanning 62 FFT bins. Yet 0.305s is significantly too long to be able to respond to the beat in real time. The linear output nature of the FFT leads to both high computational complexity and high latency in this application.

A Short-time Fourier Transform (STFT) is one approach for real-time FFT applications, but requires more computation [5]. Output can be obtained more frequently than a normal FFT where a full window of samples needs to be received, but the window function and entire transform need to be calculated every time, effectively limiting the feasible update rate in a real-time system. In addition, the STFT requires the same long window, and as such cannot avoid the perceived latency from old samples continuing to affect the output after they are no longer musically relevant.

Window functions are commonly used to reduce the size of sidelobes (noise), but all common window functions also widen the main signal peak in frequency space, degrading frequency resolution. They are also a computational overhead that scales with the update rate, since most require all input samples to be multiplied by the window at every update interval. Meanwhile, a rectangular window is essentially equivalent to not using a window function, resulting in the narrowest central peak and the highest sidelobe noise level among commonly used window functions.

I. Kanatov et al. proposed the NC method which post-processes linearly-spaced DFT outputs and provides a narrower bandwidth response than a rectangular window function can. Moreover, the NC method effectively eliminates all sidelobes and out-of-band noise without the computational overhead of window functions [4], [6]. Consequently, the window size can be reduced while achieving equivalent bandwidth to a rectangular window, meaning that temporal resolution is improved without sacrificing frequency resolution. On the other hand, the NC method is non-reversible, i.e., an inverse transform cannot be performed. That said, the NC method is applicable to the many applications that do not require an inverse transform.

The NC method is applied to linearly spaced bins, such as those obtained from a traditional FFT, in [4], [6]. However, the authors did not explain why their post-processing works or what the constraints are. To apply the NC method to non-linearly spaced FFT bins, we must first investigate the underlying mechanism and generalize the NC method. Additionally, for real-time processing, we apply the NC method to a sliding window transform, which requires specific considerations in post-processing. A sliding window transform is an extreme version of the STFT, where output bins are calculated at every input sample [7].

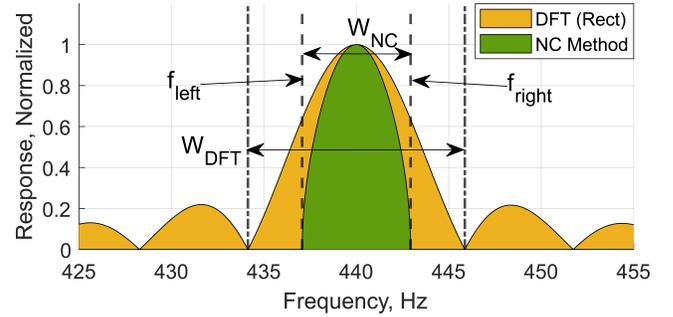

Fig. 1. Comparison of responses of 440Hz DFT and NC bins at the same window size to input signals of varying frequencies. We highlight both the narrower main lobe and the extremely low magnitude of response outside this region (noise) when using the NC method.

## III. OUR GENERALIZATION OF THE NC METHOD

Given that a regular DFT outputs a series of linearly-spaced frequency response bins (hereafter called simply *bins*), we define a single *NC bin* located between 2 DFT bins as the result of using the adjacent lower frequency ($f_{left}$) and higher frequency ($f_{right}$) bins (see Fig. 1). The *frequency* of a bin is defined as the frequency in the center of a bin's response. Then, the output of the NC bin is calculated using the real and imaginary components of the left and right bin sums [4], [6]:

$$\max(0, -(Re_L \times Re_R + Im_L \times Im_R)). \quad (3)$$

By applying the previously outlined concepts, we are able to further characterize and generalize the NC method. Neighbouring bins which are output by a typical FFT with a given window size maintain the following relationship: the left bin contains exactly 1 fewer period within that window than the right bin [8], meaning there is 360 degrees of phase shift. With window size $N$, and sample rate $F_S$, this can be expressed as:

$$\frac{Nf_{left}}{F_S} + 1 = \frac{Nf_{right}}{F_S}. \quad (4)$$

Our interpretation is that (3) uses this relationship to respond specifically only to signals that fall within the 360-degree range between the two adjacent bins that comprise one NC bin. It is required that both the left and right bins have the same window size.

Hence, we obtain a generalized formula that defines the relationship of bin center frequencies of the left and right bins that form one NC bin at center frequency $f_{center}$ by applying $2 \cdot f_{center} = f_{left} + f_{right}$ to (4):

$$\begin{aligned} f_{left} &= f_{center} - F_S/2N, \\ f_{right} &= f_{center} + F_S/2N. \end{aligned} \quad (5)$$

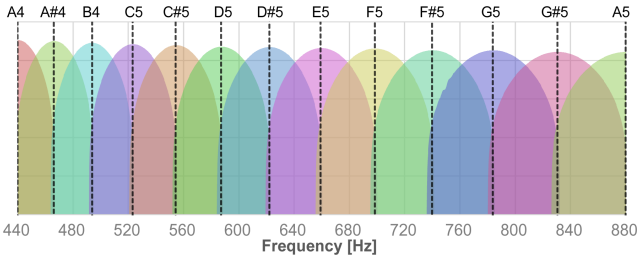

Fig. 2. The response of each NC bin in an octave, spaced exponentially to land on note frequencies. Sampling window $N$ is tuned for each note frequency.

As the frequency range to which an NC bin responds to is fully contained within the center frequencies of the two component bins, $f_{left}$ and $f_{right}$, the bandwidth of an NC bin can be calculated using:

$$W_{NC} = f_{right} - f_{left} = F_S/N. \quad (6)$$

$W_{NC}$ is half the bandwidth of the central lobe of a rectangular window DFT, i.e. $W_{DFT} = 2W_{NC}$, as visible in Fig. 1.

A key finding in this work is that for any two adjacent bins that have the same window size and obey the relation in (4), an NC bin can be created, meaning that the NC bins need not be aligned as in conventional FFTs. This means a DFT can be constructed where each resulting NC bin has a freely configurable center frequency and bandwidth. This is highly applicable to music analysis, due to the notes of interest being represented by the exponential series obtained from (1). However, $f_{left}$, $f_{right}$ and $N$ cannot be determined independently, which is discussed below.

We observe that the relationship in (6) only holds true when the window size is such that the left and right bins contain an integer multiple of half-periods of their frequencies. In effect, this means that the window size for an NC bin $k$ should be calculated from the desired $f_{center}$ and $W_{NC}$ using this formula:

$$N_k = \text{round}\left(\text{round}\left(\frac{2f_{center,k}}{W_{NC,k}}\right) \times \frac{F_S}{2f_{center,k}}\right). \quad (7)$$

We experimentally verified that as long as this constraint is met, the relation $W_{DFT} = 2W_{NC}$ holds. However, if this condition is not met, the bandwidth deviates from the result in (6), with this bandwidth widening growing significantly worse as the window size becomes small relative to the center frequency's period. Symmetry of the response around $f_{center}$ also becomes poor. We find similar problems with the response when the window size contains 2 or fewer periods of $f_{center}$.

The bandwidths mentioned in [6] ($W_{DFT} = 1.42W_{NC}$) and [4] ($W_{DFT} = 1.5W_{NC}$) are worse than our findings ($W_{DFT} = 2W_{NC}$). We believe this discrepancy is explained by their choice of window sizes not aligning with the requirement we found in (7).

## IV. PROPOSED IMPLEMENTATION

We make use of our findings to create a DFT with exponentially spaced NC bins, with each bin's bandwidth tuned such that it overlaps slightly with adjacent NC bins to create a relatively uniform response across all frequencies, as shown in Fig. 2. More concretely, NC bins following (1) are created, with $f(i)$, $W_{NC} = f(i+1) - f(i-1)$, and choice of $N$ informed by (7). Then, a DFT that includes $f_{left}$ and $f_{right}$ as bins is performed, and the output of the NC bin is obtained by (3). The window length in terms of time is $N/F_S$, approximately $1/W_{NC}$.

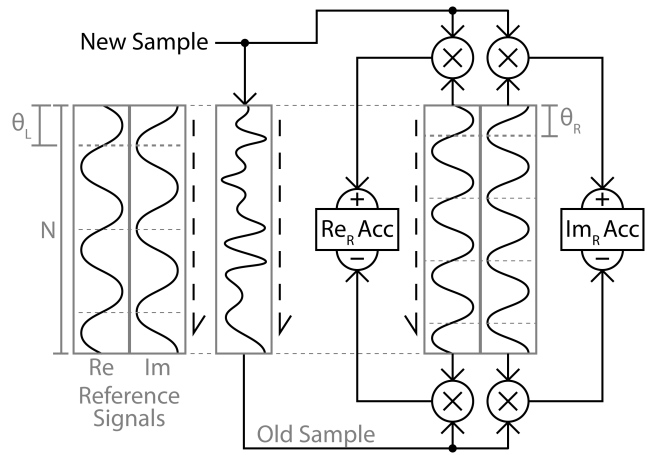

Fig. 3. Input samples are processed by both left and right bins when they enter the buffer (top), and then processed again only when exiting the buffer (bottom). The processed data is stored in the accumulators (Acc) during this time frame. Processing is shown only on the right bin, as the left operates the same aside from using reference waveforms of a different frequency. Accumulator contents are then further processed into output values.

This computation process involves complicating factors addressed hereafter in order of operations: A) The window length is too long at low frequencies, degrading temporal resolution. B) Processing audio samples efficiently requires careful implementation. C) Output data needs additional processing due to the sliding window used and the variable window length. Finally, we mention some optimizations, as well as discuss the complexity and potential disadvantages of the algorithm.

### A. Window Length

To meet responsivity goals, we target a maximum window length of approx. 0.125s, affecting only lower frequency bins, while still ensuring that the constraint in (7) is met. Otherwise, very low frequency bins could end up with window lengths so large that temporal resolution suffers, leading to an inability to distinguish beats, as well as high perceived latency as those bins continue to respond to audio that is no longer relevant. By applying this maximum window length, we employ a hybrid approach, where mid-high frequencies use constant-Q-type filtering [9], while low frequencies use variable-Q-type filtering. This is made possible by our implementation allowing for per-bin tuning of window size, which is not possible in an FFT. By using variable-Q filtering for low frequencies, specificity is reduced in these bins to maintain low latency, which is similar to a transition observed in models of the human cochlea [10].

### B. Audio Sample Processing

Our software implementation accepts audio sample data as 16-bit signed integers. When received, each input sample is multiplied by the left/right, sin/cos reference signals of the NC bin, with the results added to 4 corresponding accumulators ($Re_L$, $Im_L$, $Re_R$, $Im_R$), as shown in Fig. 3. Each sample received shifts the sliding window 1 sample forward, and likewise shifts the reference signals by $2\pi f/F_S$, incrementing their accumulated phase offsets $\theta_L$ and $\theta_R$.

Then, $N$ samples later, the same input sample is taken back out of the buffer, the same multiplication is done with the same reference signal values, and the results are subtracted from each of the 4 accumulators, netting 0. This can be thought of as an FIR filter of length $N$. All math at this stage, including reference signal computations, are done using integers to ensure that the filters remain stable over time. Floating-point computations would introduce gradual errors (i.e. a non-zero net change). Integer overflow in the accumulators and other data structures must be prevented, as it would cause irrecoverable instability.

### C. Output Data

The sliding window allows output data (NC bin magnitudes) to be calculated at any time. When this is done, it is crucial to first apply a corrective phase offset to the bin's accumulator values. This requirement is inherent to the use of a sliding window, and without this step, the left and right bins would not maintain the relative phase relationship that is critical to the functioning of the NC method. This can be expressed as a modification to (3), where $R$ is the rotation matrix:

$$R(\theta_L)\begin{bmatrix}Re_L\\Im_L\end{bmatrix}=\begin{bmatrix}Re'_L\\Im'_L\end{bmatrix}, R(\theta_R)\begin{bmatrix}Re_R\\Im_R\end{bmatrix}=\begin{bmatrix}Re'_R\\Im'_R\end{bmatrix}, \quad (8)$$
$$\max(0, -(Re'_L \times Re'_R + Im'_L \times Im'_R)).$$

To equalize each bin's response magnitude, they are divided by the respective $N$ of each bin. However, one issue remains: treble frequencies have comparatively short window sizes, which can cause missed data if the output is not used often enough. For example, the 8th octave has an NC bin with window size of just 2.6ms, much faster than a 60Hz display can show. To resolve this, we employ a simple infinite-impulse response (IIR) filter on the output values. Each NC bin's IIR filter has a time coefficient chosen to even out the response time of all bins, while keeping them highly responsive. New data is pushed into the IIR filter at time steps approximately equal to half of the minimum present window size.

### D. Buffer Optimization

We optimize memory usage by storing all NC bin's sample data in a single shared circular buffer, large enough to accommodate the longest present window size. As shown in Fig. 3, each NC bin has a limited view of this buffer according to its window size. However, because each bin's stability is reliant on the exact same sample data being present in the circular buffer during addition and subtraction from the accumulators, it is important to ensure that the buffer remains consistent even in times when the software does not receive sufficient CPU time to keep up.

### E. Computational Complexity and Limitations

Our proposed implementation of the algorithm has a computational complexity at every sample introduction linear with the number of NC bins ($O(m)$). An FFT-based implementation requires an $O(n)$ step to apply the window function, followed by $O(n \cdot \log(n))$ computations based on the window size for the FFT itself, and an $O(n)$ step to combine bins to account for the linear spacing. In the example from the Introduction, the FFT requires a window size of at least 14,634 samples to span 6 octaves (7,317 or more bins), while our implementation would require only 72 NC bins for equal minimum resolution (one bin per note). Comparing these, it becomes visible that our implementation has lower computation requirements in real-time applications which use the sliding window.

It is our understanding that the NC method we use has no inverse transform. Bin outputs also appear to become distorted if the input has multiple frequency components that fall within one NC bin. Given these major drawbacks, the proposed method is unsuitable for some applications, however it also has major advantages that make it well-suited for real-time music analysis and visualization.

### V. RESULTS

Our proposed algorithm implements independent FIR filters for each bin, with each one able to simultaneously detect signals falling into its own frequency range, while having minimal influence from frequencies outside the main lobe. Because there is no window function to apply, output can be obtained as frequently as desired, without the need for a full recomputation each time like an FFT-based STFT. This makes our proposed method highly suitable for real-time applications where output data is required many times per window, in addition to the low noise and non-linear output bin spacing. Reduced latency can be realized by reducing the window size. As the window size is proportional to $N \propto F_S/W$, rescaling the main lobe width $W$ from table I to be 1.0 for each algorithm is equivalent to scaling the window length by the same factor. This means that to achieve equivalent resolution, the window size, and thus latency, can be halved by using our proposed method as compared to a rectangular FFT.

We have developed an optimized implementation of the algorithm described in this work for the x86-64 CPU architecture in C#, running under .NET 8 on Windows 10 for testing. On a desktop PC, with an AMD Ryzen 7 5800X3D CPU, the DFT in the application took approximately 0.31 μs of time per audio sample processed when analyzing 8 octaves from 27.5 Hz to 6839.6 Hz with 192 NC bins total (24 NC bins per octave). The audio was 48 KHz sample rate stereo audio received real-time in packets of 480 samples every 10 ms from the operating system via WASAPI. The CPU frequency was observed to generally be in the range of 3.9 to 4.0 GHz.

In summary, this means that our implementation of the algorithm described in this work utilized less than 2% of one CPU core for the real-time DFT computations that encompass the entire range of frequencies of interest for music analysis, with frequency resolution generally sufficient to discern whether a

TABLE I

| Algorithm | Computational Complexity | | x86 Instructions Executed | | Main Lobe Width | Sidelobe Level (dB) |
|---|---|---|---|---|---|---|
| | Add 1 Sample | Full Window ($n$) | Add 1 Sample | Full Window | | |
| Rectangular FFT[a] | $O(n \cdot \log(n))$ | | 3.35E5 | | 1.0 | -13 |
| Hann FFT[a] | $O(n \cdot \log(n) + n)$ | | 3.84E5 | | 2.0 | -32 |
| Hamming FFT[a] | $O(n \cdot \log(n) + n)$ | | 3.84E5 | | 2.0 | -43 |
| Proposed Method[b] | $O(m)$ | $O(n \cdot m)$ | 2.58E3 | 2.11E7 | **0.5** | **None** |

$n$ denotes the number of samples, $m$ the number of bins analyzed. Sliding window transforms add 1 sample at a time, and high framerate visualizations require output after adding a number of samples much smaller than an entire window.

[a]Used FFTW [11], MSVC, with $m = 8192$, $n = 16384$. Postprocessing to reshape linearly spaced output bins omitted.

[b]Used MSVC, calling our C# DLL, $m = 144$ (6 octaves, 24 NC bins each), $n = 8192$.

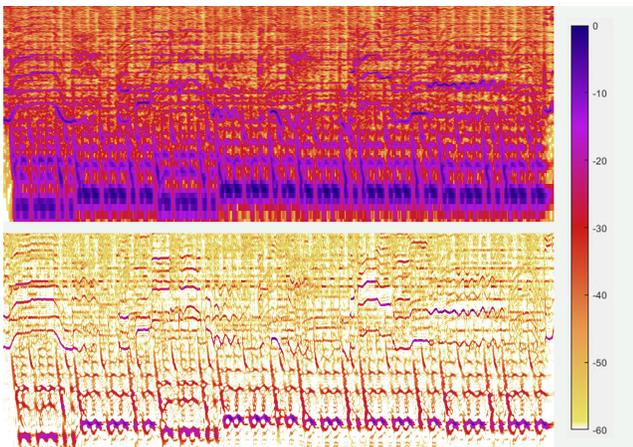

Fig. 4. The spectrogram output by a Blackman-Harris windowed FFT (top) and from our method (bottom) from an 11s excerpt of pop music. Our approach has a significantly lower noise floor, making the present notes much clearer. Unlike the FFT where bass is shown at the bottom as large low-resolution blocks, our approach is able to resolve much more detail. Vertical axis is logarithmic scale frequency from 27.5 to 3520Hz. Amplitude scale is logarithmic (dB-like), normalized to maximum response.

note was on key or not. This also translates to offline analysis being able to run at over 70x real-time on a single core.

We found 24 NC bins per octave (2 NC bins per note) to provide a good balance of speed and resolution, allowing clear distinction of notes and providing some microtonal information. On systems that are computationally constrained, 12 NC bins per octave still gives sufficient resolution to clearly distinguish notes. Fig. 4 shows a comparison between a standard FFT and our approach.

Our implementation is integrated as part of the Color-Chord.NET open-source music visualizer [12], and the code for this proposed method is publicly available there. A standalone implementation of only the DFT described in this work is also available as a DLL ready to be used from any environment that can call regular C-style exported functions.

## VI. Conclusion

In this work, we outlined an alternative implementation of an efficient sliding window real-time DFT which minimizes sidelobes and noise, while attaining improved temporal resolution and latency via shorter window sizes. The frequency and bandwidth of each output bin is adjustable, which is highly beneficial for analysis of the non-linearly spaced notes in music. Compared to an FFT, window sizes can be halved, and 2 orders of magnitude fewer CPU instructions are required to process one sample.

## Acknowledgements

We extend our deepest gratitude to Charles Lohr, Will Murnane, Samuel Ellicott, Janidhu Somatilaka, Yannick Lamarre, and 7ERr0r for their help and inspiration. This paper was proudly 100% human-written without the use of AI tools.

## References

[1] R. De Prisco, D. Malandrino, D. Pirozzi, G. Zaccagnino, and R. Zaccagnino, "Understanding the structure of musical compositions: Is visualization an effective approach?" *Information Visualization*, vol. 16, 06 2016.

[2] T. Yoshizawa, S. Hirobayashi, and T. Misawa, "Noise reduction for periodic signals using high-resolution frequency analysis," *EURASIP Journal on Audio, Speech, and Music Processing*, vol. 2011, no. 1, p. 5, Sep 2011. [Online]. Available: https://doi.org/10.1186/1687-4722-2011-426794

[3] B. S. Gowrishankar and N. U. Bhajantri, "An exhaustive review of automatic music transcription techniques: Survey of music transcription techniques," in *2016 International Conference on Signal Processing, Communication, Power and Embedded System (SCOPES)*, 2016, pp. 140–152.

[4] I. Kanatov, D. Butusov, A. Sinitca, V. Gulvanskii, and D. Kaplun, "One technique to enhance the resolution of discrete fourier transform," *Electronics*, vol. 3, p. 330, 03 2019.

[5] O. Özhan, *Short-Time-Fourier Transform*. Cham: Springer International Publishing, 2022, pp. 441–464. [Online]. Available: https://doi.org/10.1007/978-3-030-98846-3

[6] I. I. Kanatov, V. V. Gul'vansky, and D. I. Kaplun, "Method of decrease of discrete fourier transform sidelobes without window functions," in *2018 7th Mediterranean Conference on Embedded Computing (MECO)*, 2018, pp. 1–4.

[7] E. Jacobsen and R. Lyons, "The sliding dft," *IEEE Signal Processing Magazine*, vol. 20, no. 2, pp. 74–80, 2003.

[8] J. O. Smith, *Mathematics of the Discrete Fourier Transform (DFT)*. http://ccrma.stanford.edu/~jos/mdft/Orthogonality_Sinusoids.html, accessed 2025-03-07, online book, 2007 edition.

[9] R. A. Dobre and C. Negrescu, "Automatic music transcription software based on constant q transform," in *2016 8th International Conference on Electronics, Computers and Artificial Intelligence (ECAI)*, 2016, pp. 1–4.

[10] J. Leschke, G. Rodriguez Orellana, C. A. Shera, and A. J. Oxenham, "Auditory filter shapes derived from forward and simultaneous masking at low frequencies: Implications for human cochlear tuning," *Hear Res*, vol. 420, p. 108500, Mar. 2022.

[11] M. Frigo and S. G. Johnson, "The design and implementation of FFTW3," *Proceedings of the IEEE*, vol. 93, no. 2, pp. 216–231, 2005, special issue on "Program Generation, Optimization, and Platform Adaptation".

[12] C. Biesinger. (2024, Aug.) *ColorChord.NET*. [Online]. Available: https://github.com/CaiB/ColorChord.NET